# Flight-Scope: microscopy with microfluidics in microgravity

**Authors:** Thomas Wareing*[1], Alexander Stokes*[1], Katrina Crompton*[2], Koren Murphy*[1], Jack Dawson[1], Yusuf Furkan Ugurluoglu[1], Connor Richardson[1], Hongquan Li[3], Manu Prakash[3], Adam J. M. Wollman†[2]

*These authors contributed equally, †Corresponding author*

**Affiliations**

[1]*School of Engineering, Newcastle University, Newcastle-Upon-Tyne, NE2 4HH, UK*

[2]*Newcastle University Biosciences Institute, Newcastle-Upon-Tyne, NE2 4HH, UK*

[3]*Stanford University, Stanford, CA 94305, United States*

**Abstract**

With the European Space Agency (ESA) and NASA working to return humans to the moon and onwards to Mars, it has never been more important to study the impact of altered gravity conditions on biological organisms. These include astronauts but also useful micro-organisms they may bring with them to produce food, medicine, and other useful compounds by synthetic biology. Parabolic flights are one of the most accessible microgravity research platforms but present their own challenges: relatively short periods of altered gravity (~20s) and aircraft vibration. Live-imaging is necessary in these altered-gravity conditions to readout any real-time phenotypes. Here we present Flight-Scope, a new microscopy and microfluidics platform to study dynamic cellular processes during the short, altered gravity periods on parabolic flights. We demonstrated Flight-Scope's capability by performing live and dynamic imaging of fluorescent glucose uptake by yeast, *S. cerevisiae,* on board an ESA parabolic flight. Flight-Scope operated well in this challenging environment, opening the way for future microgravity experiments on biological organisms.

**Introduction**

Understanding the impact of altered gravity on living organisms has never been more important. Planned future missions to land humans back on the moon, and Mars, will require extended periods of time in altered gravity. It would take at least 7-9 months to travel to Mars. [1] This time in microgravity has an adverse effect on astronaut health, including well characterised impacts on bone density [1] but also less explored impacts on molecular and cellular processes in the immune system [[2]–[6]] and metabolism [[3], [5]–[12]]. Extended durations in space will also require new technologies in self-sufficiency and miniaturisation due to the cost of carrying cargo into space. Synthetic biology, where micro-organisms are harnessed to grow foods and medicines, is an attractive technology for space travel as it requires only minimal starting material to be carried which can be scaled up in bioreactors. Baker's yeast is a key organism for synthetic biology. Yeast has been engineered to produce medicines: morphine, [13] penicillin, [14] and breviscapine, a drug for the treatment of angina [15]. Yeast has also been used to produce haemoglobin for cultured meat [16].

Additionally, yeast is a key model organism for studying basic biological processes also happening in human cells [17].

There are many different platforms for simulated and real microgravity that can be used for research. The International Space Station (ISS) offers the longest period of reduced gravity but is difficult to access and costly. Drop towers allow the least amount of microgravity experiment time with an average of only 6-10 seconds [[18], [19]]. Sounding rockets allow 3-15 minutes of microgravity [20] but are also costly [21]. Neither of these microgravity platforms allow for operators to perform their experiments directly. Parabolic flights in specially fitted jet aircraft offer 15-20 seconds of microgravity per parabola and are therefore one of the most accessible and cost-effective microgravity platforms currently. Parabolic flights also allow for dynamic experiments whereby the effects of changes in effective gravity can be observed in real time and possible adjustments made accordingly.

The microscope has been a core tool for biology research for over 300 years [22], including in microgravity research (summarised in Table 1). Several microscopes have been used in altered gravity environments, including for permanent use on the International Space Station (ISS). These include the Bioserve Microscopy Platform (BSMP) [23], the Light Microscopy Module (LMM) [4], the Nanoracks Microscope 3 and the Mochii ISS National Laboratory (NL), the first scanning electron microscope capable of providing high-resolution images on the ISS [24]. The FLUMIAS-DEA fluorescence microscope has also been on board the ISS Space Tango Facility temporarily and used to image human macrophage cells [10]. Microscopes have also been used on the space shuttle, including the NIZEMI which used a commercial Zeiss microscope to study the behaviour of *Paramecium biaurelia* in response to gravity changes [2]. CODAG, a stereo long-distance microscope, equipped with high-speed CCD cameras was used to image dust particle motion and the structure of dust aggregates of sand flown onboard a space shuttle [3]. On parabolic flights, commercial brightfield microscopes have been used to image migrating immune cells [6] and, mounted on a silicone damped rack, the beating pattern of algal flagella [5]. Bespoke microscopes like the confocal laser spinning disc microscope, Fluorescence Microscopy Analysis System (FLUMIAS), and the digital holographic microscopy system, (DHM), have also been used on parabolic flights to image cytoskeletal changes in human follicular thyroid carcinoma cells [10] and mouse myoblast cells [8] respectively. The Mars Rover too has its own magnifying hand lens, MAHLI (Mars Hand Lens Imager), that is used to take microscopic images of minerals and structures in Mars rocks and soil [25]. Of these microscopes, none has incorporated a microfluidic system to allow live, real-time observations of changes to sample behaviour during changes to its chemical environment. Microfluidics is a key tool in yeast [[17], [26], [27], [4]] and synthetic biology research [[28]–[30]] and is an extremely useful technology for microgravity platforms which offer very little time to activate and record changes to biological behaviour in samples.

We developed Flight-Scope, a new microscopy and microfluidics platform specifically designed to operate in reduced or altered gravity environments. Flight-Scope is based on the open-source *Simplifying Quantitative Imaging Platform Development and Deployment* (SQUID) microscope [31], with an adapted design to be resistant to vibration and differing gravitational/inertial forces. Alongside this, a bespoke five-channel microfluidic pump system was built, designed to operate in low and high gravity, incorporating a bespoke microfluidic chip mounting system to facilitate fast sample changes during flights. We tested the system on board a European Space Agency (ESA) parabolic flight, imaging live yeast, *S. cerevisiae*, in microgravity and hypergravity while injecting fluorescent glucose. Our open-source platform will enable future live cell experiments under microgravity and, due to its rugged design, experiments in extreme environments on earth.

| | Year | 0g platform | Microscope type | Unique capabilities | Other capabilities |
|---|---|---|---|---|---|
| **NIZEMI** [2] | 1994 | Space Shuttle Columbia, IML-2 | Commercial (Zeiss) | Darkfield<br>Thermal control of samples (14-38C) | Brightfield<br>Mag. up to 40X<br>Movable xy stage<br>Movable z-axis |
| **CODAG** [3] | 1998 | STS-95 | Commercial | Stereo long-distance | |
| **LMM** [4] | 2000+ | ISS | Bespoke | Darkfield<br>Differential interface contrast<br>Dynamic light scattering<br>Full-field static light scattering<br>Spectrophotometry<br>Optical tweezers<br>Micro-rheology | Brightfield<br>Fluorescence |
| **BiozeroBZ-8000** [5] | 2006 | ZeroG Parabolic Flight | Commercial (Keyence) | 4 channel fluorescence<br>10-180X mag. | Brightfield<br>Fluorescence<br>Vibration dampening |
| **BioLab** [5] | 2008 | ISS | Bespoke | Incubator<br>Darkfield<br>Phase contrast | Brightfield<br>Up to 40X mag. |
| **ESA PFC 2008** [6] | 2009 | Zero-G Parabolic Flight | Commercial (Leica DMIL) | Incubator | Brightfield |
| **DHM-RPM** [7] | 2010 | RPM | Bespoke | Laser fluorescence | |
| **DHM-RPM-PFC** [8] | 2010 | RPM<br><br>ZeroG Parabolic Flight | Bespoke | Superior mag. and NA without vibration dampening<br>Numerical autofocusing | LED fluorescence |
| **DHM-SM** [9] | 2012 | SM, JASTEC | Bespoke | Laser fluorescence | |
| **MAHLI** [25] | 2012 | Mars Rover | Bespoke | Portable, automatic, white and UV light capabilities. | |
| **FLUMIAS** [10] | 2016 | Zero-G Parabolic Flight<br><br>TEXUS 52 Sounding Rocket | Bespoke | Confocal laser spinning disc imaging<br>Temperature-controlled fixation unit<br>4 diode lasers for fluorescence<br>Diode-pumped solid-state laser | Movable xy stage |
| **BioServe Microscopy** | 2016 | ISS | Bespoke | Near real-time experiment feedback<br>Hardware upgradeable | Brightfield |
| **Platform (BSMP)** [3] | | | | | |
| **FLUMIAS-DEA** [11] | 2018 | ISS, Space Tango Facility | Bespoke | 4 colours for LED fluorescence<br>Autofocus system<br>Automatic imaging with manual capabilities<br>3D imaging | LED fluorescence<br>Movable xy stage |
| **Image acquisition module** [32] | 2020 | RPM-NASA MSSF | Commercial (Dino-Lite) | Live streaming via WIFI<br>Darkfield | Brightfield<br>LED fluorescence |
| **MochiiISS-NL** [24] | 2020 | ISS | Commercial (Mochii) | Scanning electron microscope | |
| **Nanoracks Microscope 3** | 2020 | ISS | Commercial | USB microscope, 20-240X mag. | Movable xy stage |

***Table 1:*** *Summary of previous 0g microscopes where column 3 details the 0g platform the microscope was used on and columns 4-7 outline how the microscopes differ from the Flight-Scope.*

## Results

### Design of a vibration resistant microscope

The core of Flight-Scope is a vibration-resistant fluorescence microscope. We designed the microscope based on the open-source SQUID microscope [31] (Fig.1A). The microscope was capable of two key imaging modes - brightfield and fluorescence. These were implemented using a top-mounted RGB LED pixel matrix for brightfield and a 470nm LED for epi-fluorescence. Images were formed using a 40x/0.75 Olympus Plan Fluor objective, a f = 75 mm imaging lens as tube lens and a camera with Sony IMX178 CMOS sensor. Standard microscope slide-based samples were mounted on a bespoke 60-60mm x-y motorised stage. The microscope was controlled using a joystick and custom software [31] for image-viewing, capture and control of x-y-z motion. The microscope has a field of view of 467 um x 311 um (3072 x 2048 pixels) with sample side pixel size of 152nm. At full resolution the camera supports 60fps. The software allowed switching between brightfield and fluorescence channels during imaging at 0.5 frames/second. Fluorescence illumination intensity could reach 0.05-1mW/mm$^2$. Vibration resistance was engineered into the microscope by mounting the optical train on a 2" dampened post (Fig.1Aii) and the experimental setup placed on four dampeners.

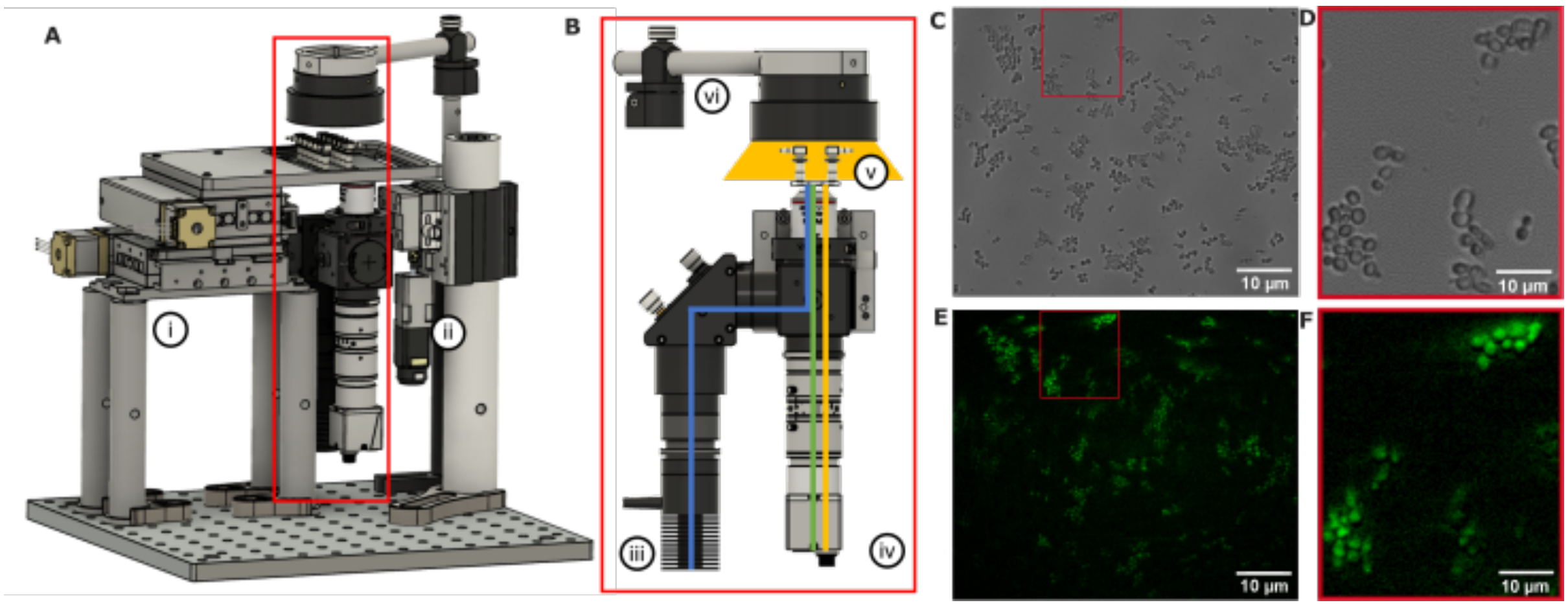

***Figure 1 A.*** *Flight-Scope microscope with (i) motorised xy stage (ii) a central vibration mast which absorbs vibrations felt by the optical train.* ***B.*** *The microscope optical train is shown in the red outline and consists of (iii) a fluorescence LED, (vi)*

*brightfield LED array, (v) sample and (iv) camera. The blue and green lines depicts the excitation and emission optical paths respectively. The yellow line depicts the brightfield optical path.* ***C.*** *A brightfield image of yeast taken with Flight-Scope with* ***D.*** *showing a zoomed in region.* ***E.*** *A fluorescence image of yeast stained with fluorescent glucose 2-NBDG.with* ***F.*** *showing a zoomed in region.(100ms exposure, 20% illumination power)*

We tested the imaging capability of the microscope by imaging live yeast, *S. cerevisiae,* in brightfield (Fig.1C-D) and in fluorescence (with fluorescent glucose analog 2-NDBG, Fig.1E-F). We also characterised the vibration resistance of our microscope on the ground by mounting the microscope on to a tuneable vibration source (centrifuge) and imaging live yeast cells (Fig.2A-B), reaching vibrations of up to 13000rpm (217Hz), similar to the range of vibration onboard a flight [33]. Vibration in the image was quantified by calculating the cross-correlation between consecutive images in MATLAB. We found that we could image yeast cells successfully in these conditions and the correlation only dropped to a minimum of 0.6 (Fig.2C). We also tested the vibration resistance of the microscope by placing it inside a vehicle – a van - (Fig.2D-E) with the engine running idly and found greater but tolerable changes in the cross-correlation of the yeast image videos (Fig.2F).

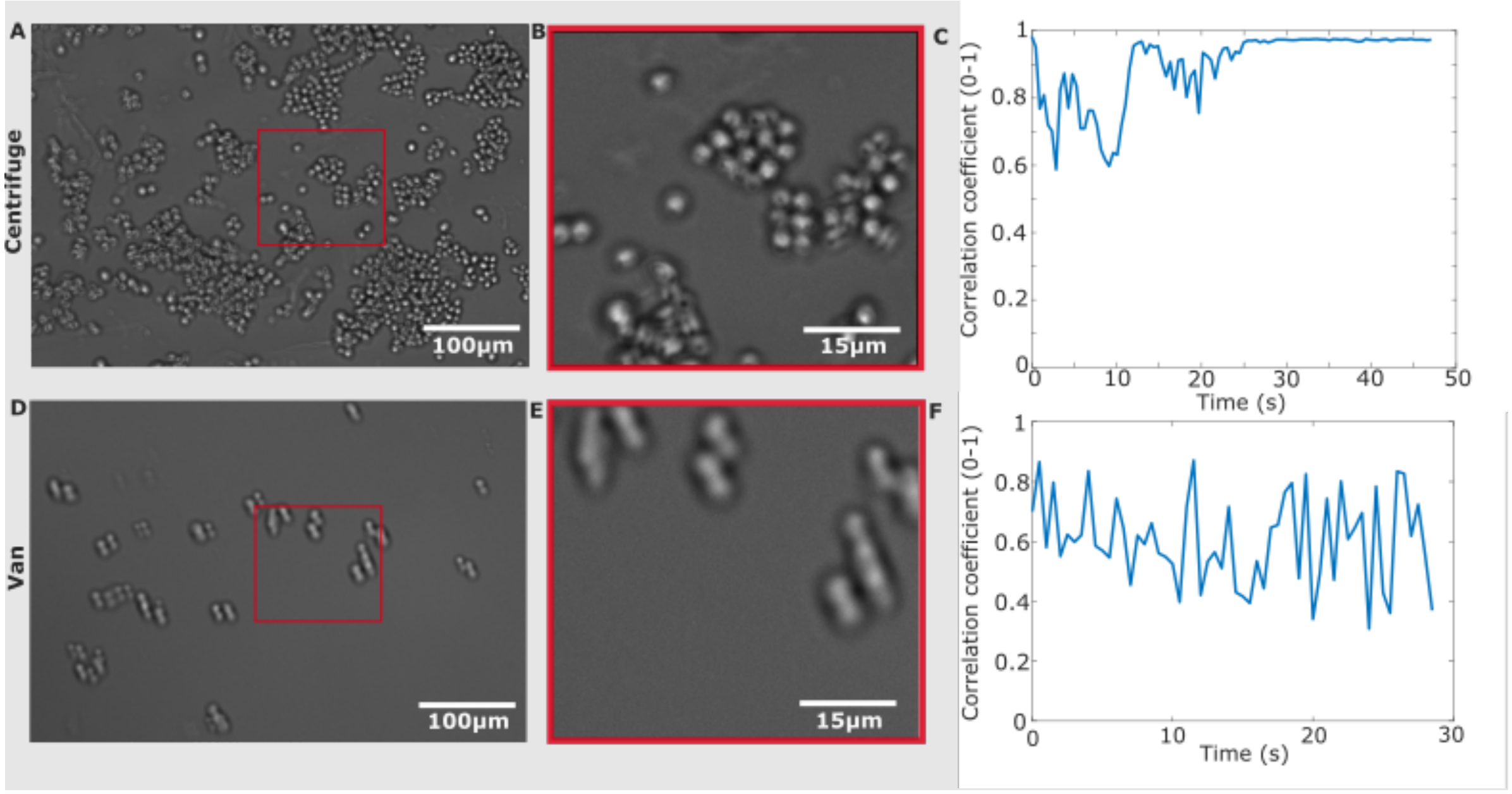

***Figure 2 A.****Brightfield image of yeast cells taken while Flight-Scope was mounted on a benchtop centrifuge.* ***B.*** *A zoomed-in ROI brightfield taken on the centrifuge.* ***C.*** *Plot of correlation coefficient of cell position in consecutive frames vs time for images taken while the microscope was on the centrifuge.* ***D.*** *Brightfield image of yeast cells taken while the microscope was mounted inside a van.* ***E.*** *A zoomed-in ROI brightfield taken on the van.* ***F.*** *Plot of correlation coefficient of cell position in consecutive frames vs time for images taken while the microscope was mounted inside a van.*

## Microfluidic system

We designed a bespoke microfluidic system to allow dynamic changes of the sample fluidic environment during the short gravity changes on board a parabolic flight (Fig.S2A-B), designed to operate in altered gravity and with aircraft vibration. The system used stepper motors on a 3D-printed chassis (Fig.3A-B) to drive 5 syringes with flow rates of 75µL/s using 1mL syringes. The system was controlled with a custom user interface (Fig.3A(i),(ii)) that allowed syringe selection, dosing, jogging forward and backward and homing (Fig.S4A) pumps for fresh syringes. All 5 syringes could be changed over, thanks to an innovative clasp-based panel design (schematic in Fig.3C and shown mounted in Fig.3A(i)), with all 5 syringes mounted to a panel and connected to a microfluidic chip.

We tested the system by imaging the dynamic real-time uptake of fluorescent glucose in yeast cells. We adhered cells using lectin, Concanavalin A, in a microfluidic chamber and pumped in fluorescent glucose analog, 2-NBDG (Fig.3D-F). Yeast cells became increasingly fluorescent as they took up glucose (Fig.3F). We quantified the fluorescence increase due to glucose uptake in each cell using bespoke MATLAB software, FRETzel [34], observing rapid cellular uptake (Fig.3D).

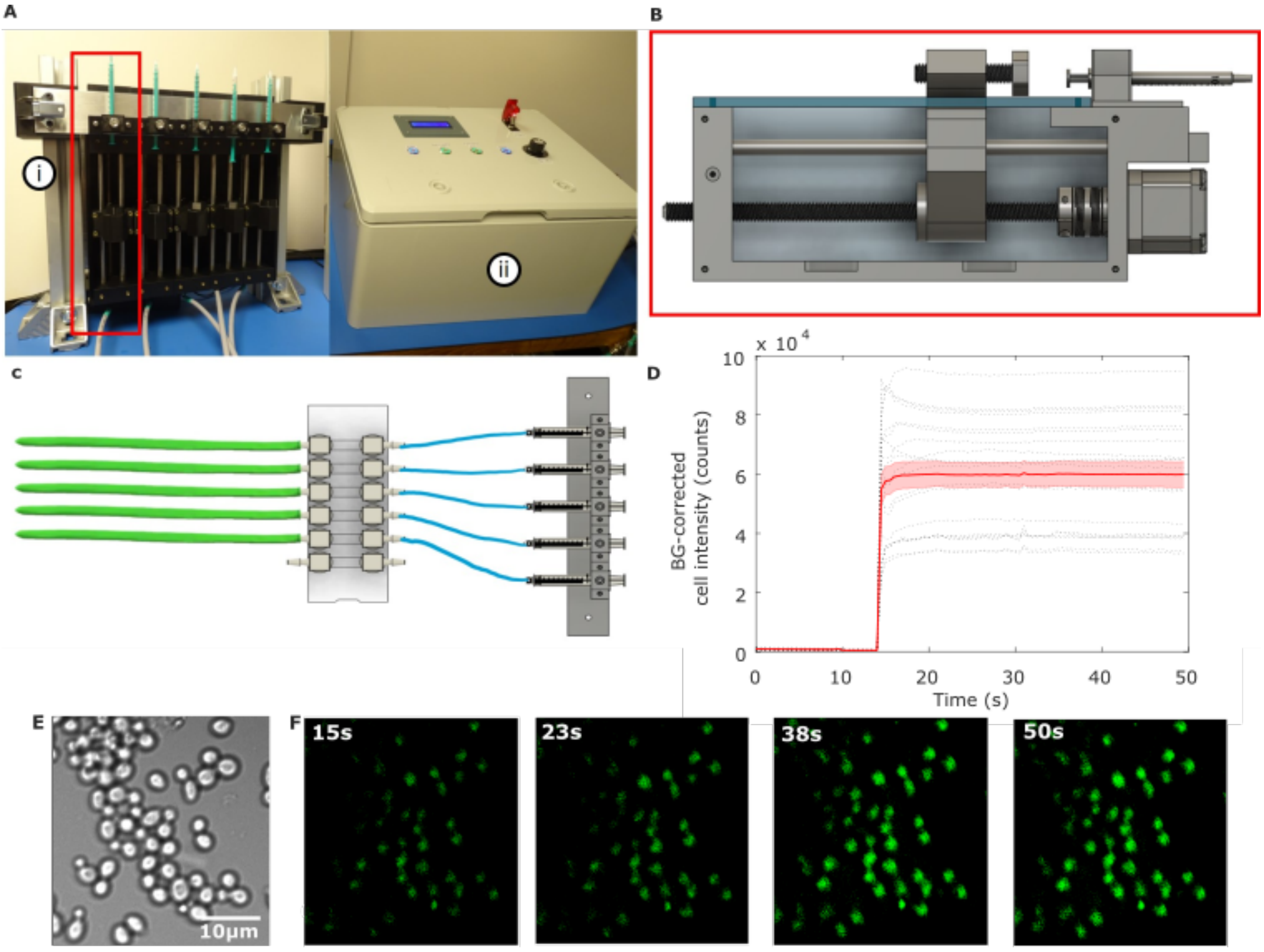

*Figure 3 A. The Flight-Scope syringe pump with clasp-based syringe rack (i) and user interface (ii).* ***B.*** *Syringe driver schematic.* ***C.*** *Panel-mounted sample system with syringe-rack, inlet ports (right) and outlet ports in green (left).* ***D.*** *Fluorescent 2-NBDG uptake vs. time of yeast cells for 10µM 2-NBDG concentration n=20 cells, black dotted lines are individual cell traces with their mean (red lined) and standard error (red shading).* ***E.*** *Brightfield image of yeast taken with the Flight-Scope and* ***F.*** *fluorescence images taken over a course of 15,23,38,50 seconds after 2-NBDG injection showing increases in cell intensity over time.*

Designing microscope for operation on a parabolic flight

We further engineered Flight-Scope for operation on-board a parabolic flight. The microscope needed to be leak-proof to prevent substances coming in to contact with experimenters during the microgravity portions of the flight. The system was encased in a protective and leak-proof Zarges box (Fig.4). The system also had to be mechanically constrained to prevent movement during flight and to ensure it would remain fixed in the event of a plane crash (Fig.S1B-C). Thus the system was bolted to the plane seat rails and mechanical dampeners were added to reduce vibration from the aircraft (Fig.S1C). The waterproof Zarges box lid was also interlocked with the fluorescence LED source to prevent dazzling stray light when opened.

We designed a closed microfluidic sample to image on board the parabolic flight and test the systems capability. Yeast cells were adhered to 5-channel microfluidic slides connected to syringes mounted to the syringe pump. The experiment consisted of 6 individual microfluidic sample chips, each consisting of 5 channels on a slide with each channel connected to syringes and waste ports via silicone tubing (Fig.3B). Each sample system was designed to be used once for each parabola set (Fig.S2B) (five parabolas in a parabola set, each lasting 100s, separated by 90s and 5 mins between sets (Fig.S2A-B). Thus, a 3D printed sample holder system was designed to facilitate fast changeovers during parabola set breaks (Fig.4(iii),(iv),(v)). A changeover requires one operator to home the syringes while the other operator unclips the old sample rack from the pump, secures it in the holder, before removing and attaching a new sample rack (Fig.S4A).We found that our sample holder system allowed a changeover to be performed in under 3 mins, allowing enough time to close the lid ready for the next parabola set (Fig.S1A).

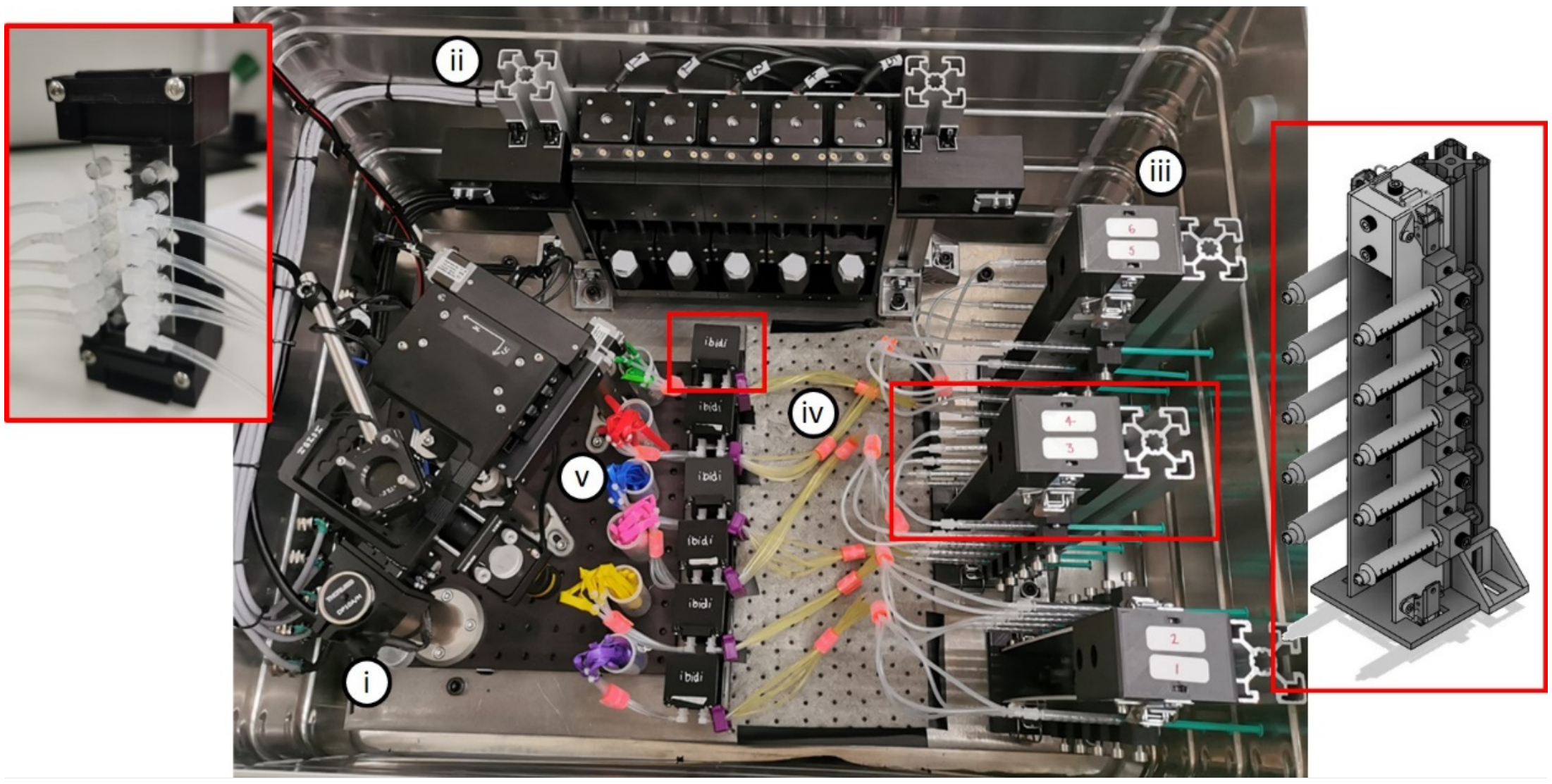

***Figure 4*** *(i) Flight-Scope and (ii) syringe pump inside the Zarges box with (iii) three syringe rack holders, also depicted in the rightmost red outline, each holding a sample set, depicted in the leftmost red outline with (iv) primed glucose inlet tubing and (v) stowed outlet balloons.*

## Performance of Flight-Scope on parabolic flights

Despite the challenges of operating on board a parabolic flight, Flight-Scope performed well. We were able to obtain similar quality brightfield and fluorescence images of live yeast during the flight (Fig.5A) as were obtainable on the ground (Fig.1C-F). We also characterised Flight-Scope's tolerance to vibration on board the flight, as before, by quantifying the correlation coefficient between consecutive images (Fig.5C). Vibration tolerance was similar to our ground tests, with correlation coefficients remaining at between 0.5-0.8 during steady state (non-parabolic) flight (Fig.5C), comparable to ground performance under similar vibration (Fig.1C,F).

During parabolas and changes in apparent gravity, to microgravity and hypergravity, we detected an xy lateral shift of 8.3±0.3$\mu$m at the sample plane. This was easily corrected for during analysis. We performed glucose uptake experiments with 2-NDBG and live yeast as on the ground. We obtained proof-of-principal data of glucose uptake during the shift from microgravity into hypergravity (0.01-1.8g). The intensity of the fluorescence images taken during 0.01-1.8g flight was characterised for 0.01-1.8g flight against that for images taken on the ground and plotted against time post 2-NBDG injection (Fig.5D). These were fitted with exponential uptake fits [34] with characteristic uptake times of 0.6 +/- 0.2 compared to 1.3 +/- 0.3 on the ground. These measurements serve as important proof-of-principal of Flight-Scope, the need to study cell signal transduction in microgravity and the potential for further microgravity experiments going forward.

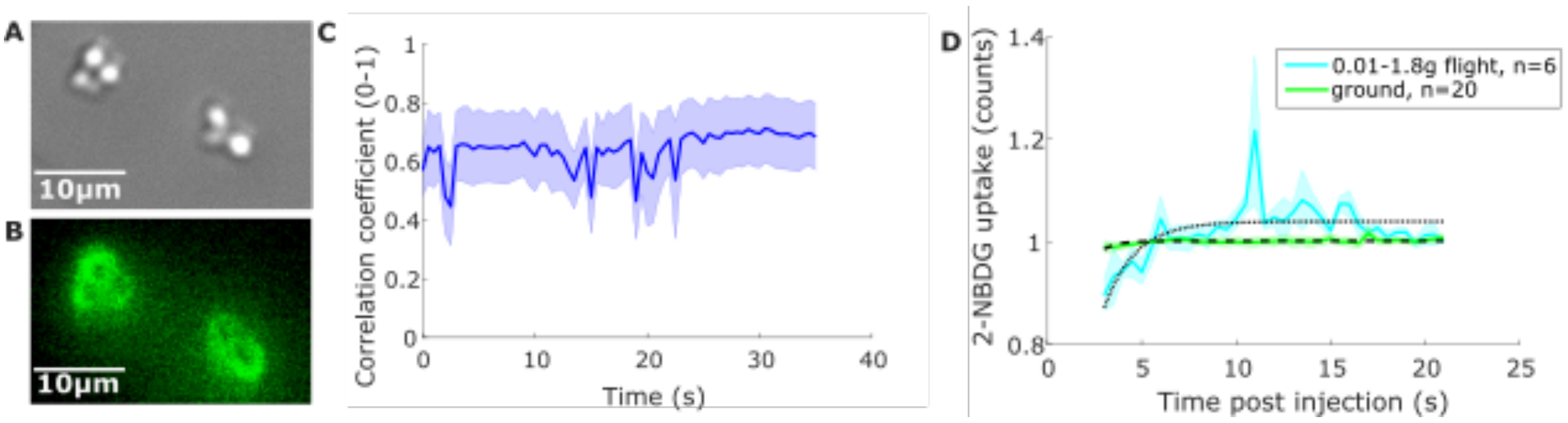

***Figure 5 A.*** *Brightfield image of yeast cells and* ***B.*** *fluorescence image of 2-NBDG uptake by the cells taken by Flight-Scope on board a parabolic flight.* ***C*** *Plot of average lateral cell location correlation coefficient for four videos taken in steady 1g flight.* ***D.*** *Plot of 2-NBDG uptake by cells when in 0.01-1.8g flight (cyan) and on the ground (green) with lines of best fit.*

## Discussion

Flight-Scope is a new microscopy and microfluidic platform for microgravity research. We demonstrated Flight-Scope's capabilities on board an ESA parabolic flight imaging live yeast cells and using our microfluidic system to expose cells to fluorescent glucose in real time during the gravity changes on the flight. The microscope hardware and software performed well and allowed both brightfield and fluorescence imaging, with minimal influence from vibration, as quantified by cell correlation coefficient which remained between 0.5 and 0.8 (Fig.5D). We did observe a lateral displacement in xy when moving into altered gravity conditions on the flight of ~8µm, although we could easily correct for this in our analysis. Our microfluidic system performed well on the flight, enabling us to inject fluorescent glucose and observe glucose uptake during changes from micro to hypergravity. Although our flight time did not allow a full program of experiments investigating how cell signalling is impaired in mircrogravity, our proof-of-principal results with Flight-Scope open the way for future investigations.

Compared to the other sixteen microscopes used in microgravity and hypergravity environments that we have identified and outlined in Table 1, Flight-Scope has a unique feature set, optimised for life science research on a parabolic flight, including brightfield and fluorescence microscopy, as well as fully motorised xyz control. FLUMIAS-DEA is one of the few microscopes, like Flight-Scope, purpose built for low gravity operation and boasts some of these capabilities [11] but lacks brightfield and microfluidics. None of the other microgravity microscopes to our knowledge have microfluidic capabilities. Microfluidics enable the kind of dynamic experiment we demonstrate here, where cell signalling can be observed in real time during changes in gravity. We demonstrate this imaging glucose uptake of budding yeast but Flight-Scope could be used to investigate many other cell signalling processes in microgravity, such as insulin signalling in human cells which has shown to be impaired [35] or cancer cell response to chemotherapy, also shown to be modulated by microgravity [36].

Future iterations of Flight-Scope could benefit from a high speed autofocus or self-stabilising system to correct for the drift we encountered imaging on the plane. The most recent SQUID platform supports autofocus using a contrast based method or a laser autofocus. A laser autofocus system could be implemented in Flight-Scope to track the coverslip. Laser-based autofocus works well with dimmer samples and is faster than the traditional image-based autofocus [37]. The laser would not pose a risk of bedazzlement due to the containment of the microscope. Alternatively, autofocusing/stabilisation can be achieved with image based methods such as with an electrically-tuneable lens (ETL) [38]. These ETLs are liquid lenses with a motor bobbin controlling the lens volume to focus the images. Although studies into the effects of gravity changes on the liquid within the lens would be required for implementation in Flight-Scope. Rapid Autofocus via Pupil-split Image phase Detection (RAPID) autofocuses the image using phase detection and could also be implemented on Flight-Scope [39].

In total, the cost of Flight-Scope was below £10k (GBP) with the microscope costing ~£6k, housing and fixings ~£2k and the 5 channel syringe pump system only costing ~£800 (GBP). This makes Flight-Scope an extremely cost-effective instrument. This was in part facilitated by 3D printing which enabled rapid prototyping of syringe pump components to accelerate design and production of Flight-Scope. Of the sixteen microscopes outlined in Table 1, only the Image acquisition module [32] is known to also have 3D printed components in its design.

Finally, Flight-Scope lends itself to a multitude of other use cases; particularly unusual environmental conditions. The system is fully contained, keeping samples safe from liquids, dust and wind. Therefore, it could be used in harsh conditions such as deserts and jungles. It is readily transportable in a vehicle with the dampening provided to the microscope ensuring that it can function with travel vibrations. Flight-Scope is a versatile platform for microscopy and microfluidics in microgravity and other extreme environments.

## Materials and Methods

### Microscope design

The main optical train (Fig. 1B) of the microscope includes a 60fps Daheng Imaging MER2-630-60U3M USB camera with Sony IMX178 sensor (2.4µm pixel size), a Daheng Imaging 75mm lens tube, an Olympus UPLFLN 40X air objective with focal length 180mm, 0.49µm depth of field and 0.75 NA (numerical aperture), a Thorlabs 809mW 470nm LED used for fluorescence illumination and an Adafruit DotStar high density 8x8 RGB LED pixel matrix grid used for brightfield illumination. The optical train, including the camera, lens tube, fluorescence light source and objective, was mounted to a dampened pillar, while the brightfield matrix was mounted on a separate adjustable height pillar, on a hinged arm to allow access (complete list of parts available in the supplementary table 1). Focusing was achieved by movement of the entire optical train in the z-axis by a Nema 8 linearly actuated motor and rail with return spring. A motorized 60x60mm xy stage was mounted separately with 4 pillars to an optical bread-board along with the optical train and brightfield pillars. A console with buttons, rotary nob and joystick is used to control the z-axis and xy stage respectively (Fig.S3B(iv)), and a printed circuit board (PCB) for control of the lights and movement through a dedicated laptop to display the microscope's image and control movements through a custom-built Python-based software, on a laptop running Linux allowing the user to change brightfield and fluorescent illumination intensity , exposure time and video recording length as well as control the motorised stage (Fig.S3A). The instructions on Github outline the installation of both the software and the drivers to the camera and the control plates, available form https://github.com/hongquanli/octopi-research

### Microfluidic system and storage

To deliver the 2-NDBG to the yeast during imaging, a bespoke microfluidic system was designed (Fig. 3A-B,4(ii)). Microfluidic sample systems consisted of 1mL syringes connected by female and male luer lock connectors, in turn connected to 6-channel microfluidic chip slides (Ibidi µ-Slide VI 0.4, Germany) with silicone tubing (2mm inner diameter, 3mm outer diameter). For waste let out of the slides following glucose injection, a balloon cable-tied around the silicone tubing was used (Fig.3C).

During the flight campaign, there were a total of 30 microfluidic sample systems, one for each parabola during the flight, in the form of 6 sets of systems. Each system consisting of 5 syringes connected, via the silicone tubing, to 5 channels in a sample slide. All samples needed to be secured to prevent the syringes from prematurely injecting during the changes in gravity. The storage system for the syringe panels used a locking mechanism; panels were stored vertically with the syringe plungers pointed towards the wall of the box (to prevent experimenters accidentally knocking them) (Fig.4(iii)) and right red frame). Each storage unit can hold two panels. The units have a 40x40 extrusion frame with 3D printed part attached. The slides were stored vertically inside a protective 3D-printed housing (Fig.4 left red frame); these were attached to the floor of the Zarges box with Dual Lock with space for the tubes between the slide holder and syringe storage unit (Fig.4).

The syringe pumps are intended to be used in a bank of five pumps (although the system can be adapted to handle more). The experiment demanded the syringe pumps be independently controlled as a fresh syringe was required for each parabola (for 6 sets of 5 parabolas). The five syringes are attached to a panel so all five syringes can be removed at once (Fig.3C), this was necessary to meet the short changeover time of 5 minutes between parabola sets.

Syringe pump design

The syringe pump was designed with modularity for easy part replacement. Each pump module (Fig.3B) facilitates substitution in case of failures. The design accommodates the addition of extra pumps, with our experiment utilizing five modules. Made from standard parts, the frame is 3D-printed, the linear motion that creates the pushing force on the syringe plunger is generated by a Nema 17 stepper motor coupled with an M10 threaded rod (chosen over a traditional lead screw to minimise cost and maximise the accessibility for build), a captive nut is concealed within the carriage, the carriage moves along two ground rods. The fine adjustments can be made to the plunger pusher by twisting the bolt to meet the plunger without prematurely injecting the syringe.

The syringe pump components were fabricated using a Flashforge Guider 2 3D printer with a 0.4mm diameter nozzle. Printing parameters were adjusted to ensure optimal printing time and quality, including a layer height of 0.2mm, four perimeters wall thickness, five solid layers at the top and bottom, and an infill density of 35%. PLA material was selected for printing as it is readily available and easy to work with.

A bespoke control system was built to independently drive the five syringe pumps used during the experiment. This system consisted of five open-loop stepper motor driver boards, five limit switches, several panel-mount inputs, and a display, all wired to an Arduino Uno flashed with a custom control script. Fig. S4E provides a simplified block diagram of the control system components and wiring. Fig. S4C provides a top-down view of the components wired together within the control enclosure. Fig. S4D shows the panel-mount connectors used to interface the control enclosure electronics with the external pump motors and limit switches, a 24Vdc power supply, and a monitoring computer/5V power supply.

As each syringe pump had to be operated independently, the stepper motors used in each pump were driven by separate open-loop driver boards. Each of these motor driver boards were provided separate STEP (sometimes referred to as PULSE) and DIRECTION signals (Fig.S4F) that were generated by an Arduino Uno flashed with a custom control script. Limit switches attached to each syringe pump were wired to this Arduino, and their states (Open or Closed) were used during individual motor homing. The state of each limit switch was also used throughout pump operation to prevent the carriage from crashing and damaging the pump in the event of misuse (overdriving pumps, for instance). The STEP and DIR signals discussed here were generated by the Arduino based on user input and the states of these limit switches.

An input panel consisting of four push buttons, a motor selector knob, and a display was provided to allow users to control the system. This control panel is shown in Fig.S4A, and the connections made to the Arduino Uno within the control enclosure (Fig.S4B).

A motor selector knob was provided to allow users to select a syringe pump for control, and four push buttons were supplied to trigger different actions: initiating motor homing, jogging, or dosing with a predefined volume of glucose (Fig.S4A). A 16x2 character LCD was included in the control panel to give the user succinct pump operation feedback, ensuring the correct operations were performed during testing. A 24V cut-off switch was provided on this control panel to enable quick power isolation in case of motor stalls or potential user injury caused by equipment misuse. Full design available here https://github.com/Alexander-Stokes/Flight-scope .

Experiment containment and plane interface

There were several constraints to experimental operation during a parabolic flight set out by Novespace, the company running the flight campaign, to ensure the safety of everyone on-board the flight.

The experiment was fixed within a 750x550x580mm Zarges box to prevent liquid leakage during flight and to supply a rigid structure and a dark space for fluorescent imaging, adhering to liquid and laser containment safety guidelines (Fig.S1B(i)). The leak-proof capability of the system was tested by filling the Zarges box with approximately 50 litres of water (prior to fitting the electronics). Further leak-proof measures included sealing the fluidic loops at both ends and covering electronics in silicone to prevent liquid causing a short circuit within the box. All electrical components were scrutinised by Novespace to check that the wire gages were thick enough to prevent burnout causing a fire risk to the aircraft, as well as checking there were sufficient current limiters and fuses within the components preventing a current surge. The experiments power came from the plane at 230V AC 50Hz and had a fuse and emergency stop at the outlet.

All equipment within the box was bolted to a 10mm aluminum base plate which was fixed to the box through four dampeners to further reduce the effect of vibrations generated by the aircraft. A further two 10mm aluminum plates (875x130mm) were bolted to the seat rails of the Airbus A300, as the box's interface with the plane (Fig.S1B(iii)) (with laptop, microscope control box and syringe-pump control interface all fitted to the baseplate with Dual Lock (Fig. S1B(ii))). This fixation received rigorous evaluation by Novespace to ensure the box was safely and securely attached to the aircraft, all components must withstand a 10G impact, the forces incurred from a worst-case scenario crash-landing.

Sample preparation

*Saccharomyces cerevisiae* CEN.PK 113-7D yeast was scraped from a YPD agar dish (10g/L yeast extract, 20g/L peptone, 20g/L agar, 4% glucose (w/v)) and cultured overnight in a 30°C shaking incubator in YNB media (13.8g/L Yeast Nitrogen Base (YNB), 4% glucose (w/v)). 200µL of the overnight culture was then pipetted into a morning culture, consisting of the same volumes of media and left in a 30°C rotating incubator for two hours. Final yeast cultures were taken from this and diluted 1:10 in fresh YNB media. 30$\mu$L of Concanavalin A was pipetted into each channel in each sample slide and left to rest for 15 minutes followed by washing with 30$\mu$L YNB. 30$\mu$L of sample culture was then pipetted into each channel and left to rest for 15 minutes. After another YNB wash, the silicone tubing filled with fluorescent glucose analogue, 10,25 or 50µM 2-NDBG (2-(N-(7-Nitrobenz-2-oxa-1,3-diazol-4-yl) Amino)-2-Deoxyglucose), syringes and waste outlets were attached.

**Acknowledgements**

The authors would like to thank the staff at the European Space Agency (ESA) Education department including Nigel Savage, Felix Scharnhoelz, Jeffery Gorrisen and Neil Melville. We would also like to thank the staff at Novespace, especially Yannick Bailhe, for ensuring our experimental rig was safe for flight. Thank you to Derek McCusker from Bordeaux University for providing us sterile water during the campaign. Thank you Annachiara Scalzone for your initial work and support for the SUGAR project. Thank you Kenny Dalgarno from the Mechanical Engineering Department at Newcastle University, UK for facilitating departmental funding. We would like to thank the Newcastle Bioimaging facility. We would like to thank our funders: the Academy of Medical Science to A.J.M.W through SBF007\100046, Wellcome Trust (grant number 204829) through Fellowship A.J.M.W. through the Centre for Future Health at the University of York, ESA Education, the UK Space Agency, Newcastle University Doctoral College Enhancement fund, Zarges and the Newcastle University Student's Union.

**Supplementary Material**

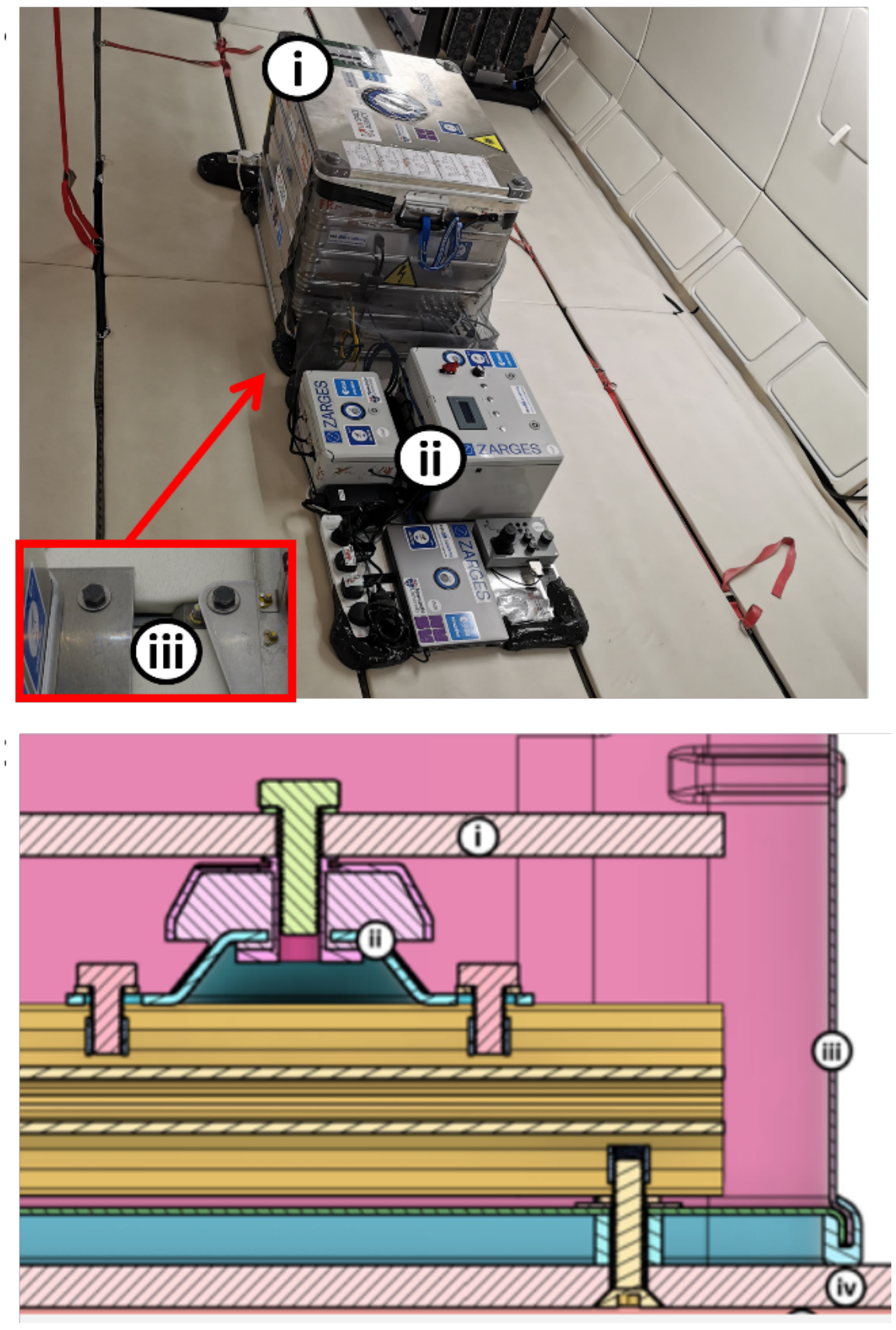

***Figure S1 A**. Top:**.** Flight-Scope Zarges box (i) and control baseplate (ii) fixed to the Airbus A300 seat-rails with bolts shown in red box (iii). Bottom:  Zarges box fixation to seat-rails cross-section (i) inner base plate – microscope and syringe pumps fix to. (ii) dampener. (iii) Zarges box. (iv) external base plate 1/2. (v) seat rails.*

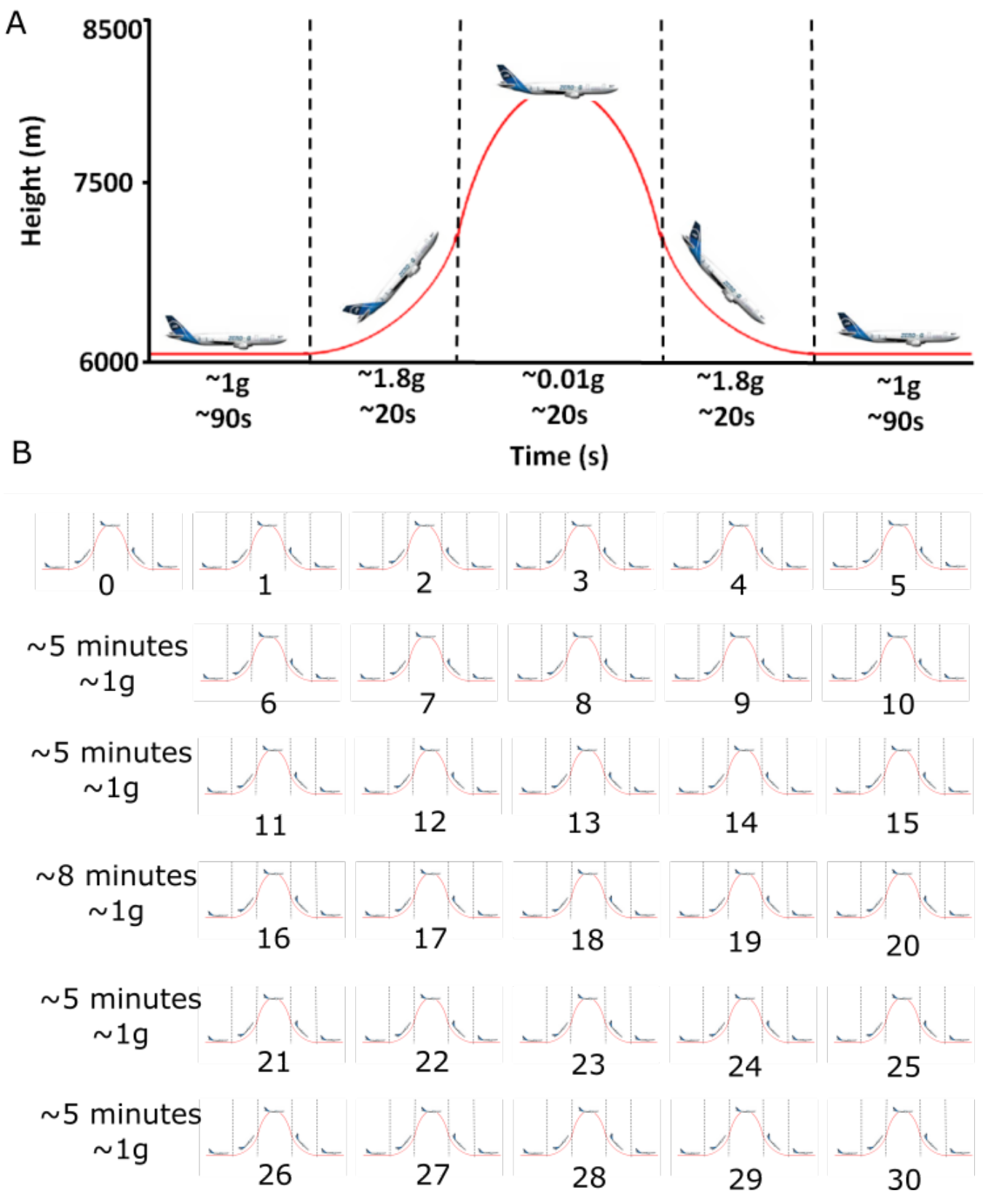

***Figure S2 A.*** *Individual parabolic manoeuvre performed during a Novespace parabolic flight indicating an approximate 20 second interval of ~0.01g sandwiched by two approximate 20 second intervals of ~1.8g. There is an approximate 90 second interval of steady flight, ~1g before and after the varying gravity. At steady flight, the plane flights at approximately 6000m but reaches up to 8500m when performing the parabolic manoeuvre.. .* ***B****. Sets of parabolas separated by approximately 5 minute or 8 minute long breaks. Each flight consisted of 31 parabolic manoeuvres taking on the profile in A.*

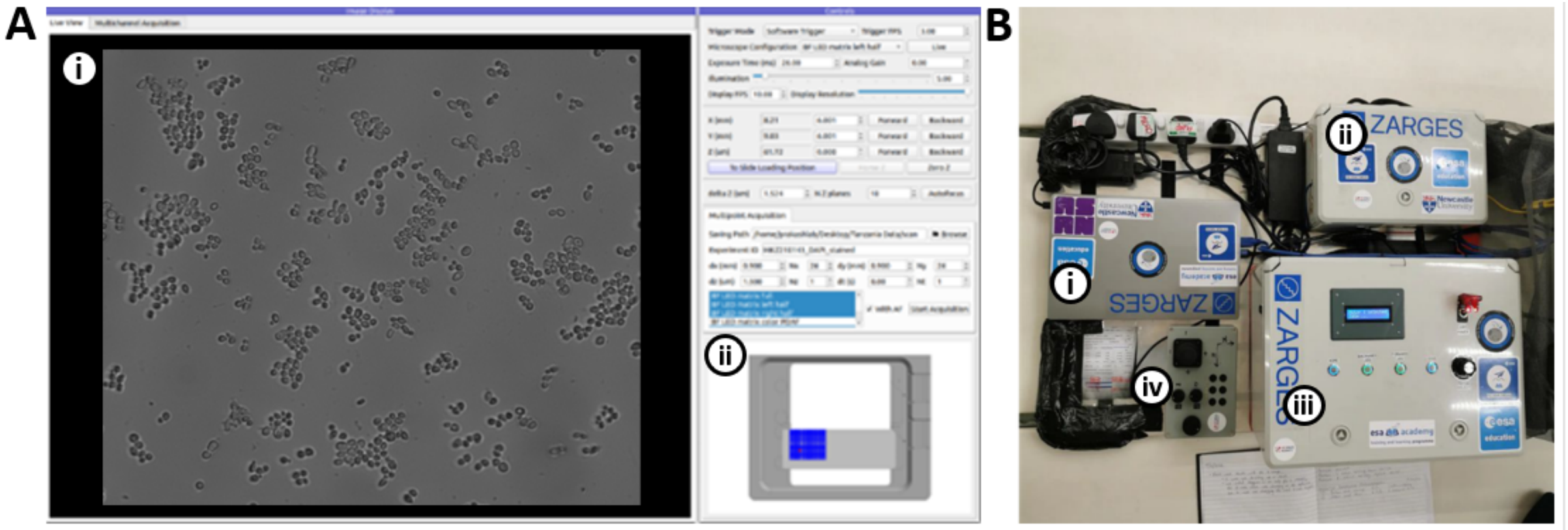

***Figure S3 A.*** *Flight-Scope user interface with as seen on the laptop. (i) field of view live feed from the microscope imaging yeast. (ii) microscope and stage computer control and settings, and map of the microscopes field of view relative to the stage.* ***B****. Control base plate layout (i) laptop. (ii) microscope PCB. (iii) syringe pump control box. (iv) xy stage joystick and z axis dial controls****.***

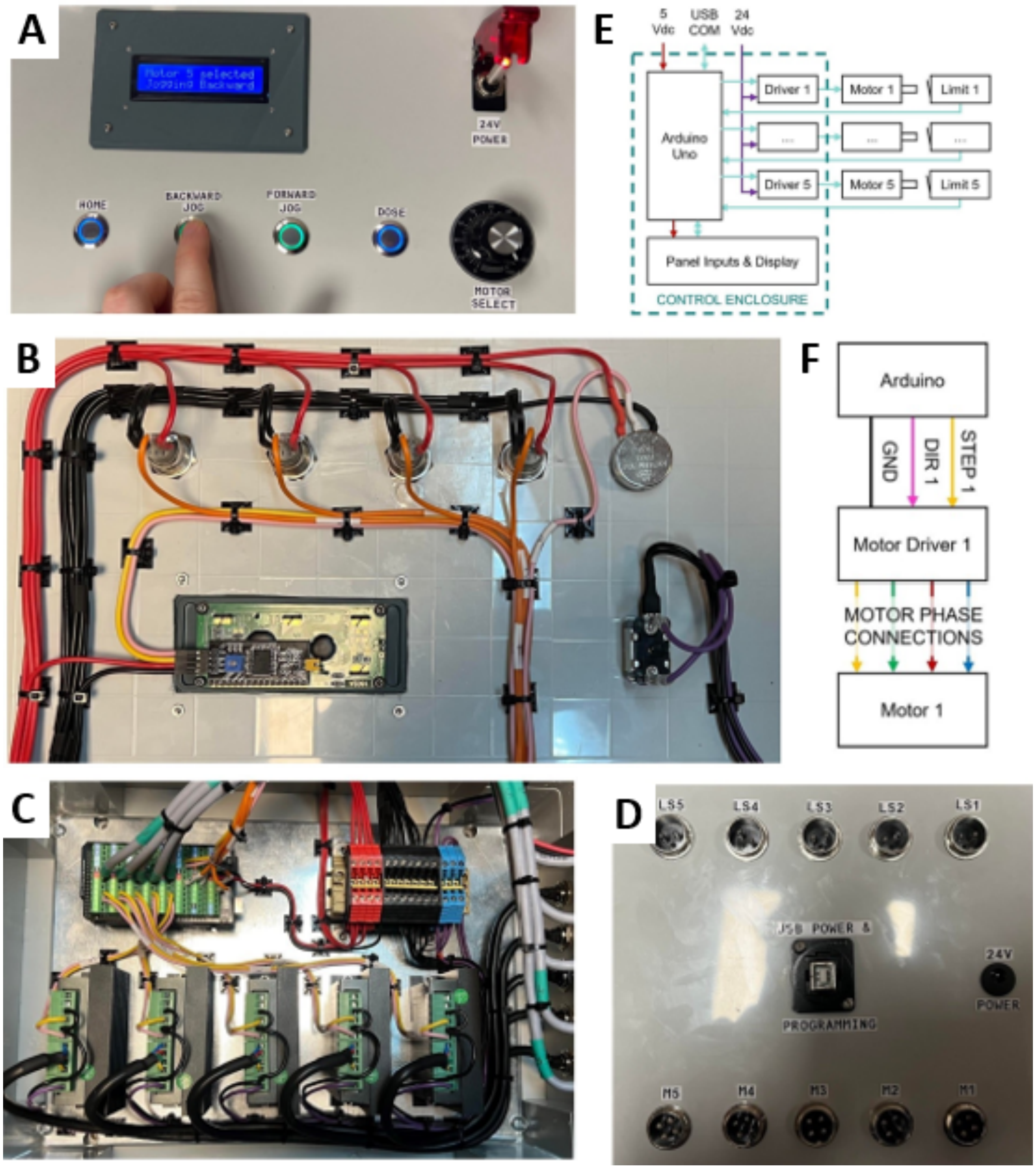

***Figure S4 A.*** *Syringe pump control panel.* ***B.*** *Wiring within the syringe pump control box.* ***C.*** *motor drivers and Arduino wired together inside the control box.* ***D.*** *motor, limit switch, power, and USB communication connections on the control box exterior.* ***E.*** *Simplified schematic of component connections and communications.* ***F.*** *simplified schematic of the connections between each motor, its controller, and the Arduino.*

Supplementary Table S1. Bill of materials

Microscope and controls

| Part | Vendor | Code | Quantity | Total cost (£) |
|---|---|---|---|---|
| 0.3" SM1 lens tube | Thorlabs | SM1L03-P5 | 1 | 44.61 |
| 0.5" SM1 lens tube | Thorlabs | SM1L05-P5 | 1 | 46.19 |
| Mounted LED - 470 nm | Thorlabs | M470L5 | 1 | 170.94 |
| Adjustable lens tube | Thorlabs | SM1V05 | 1 | 23.37 |
| Extra thick retaining ring | Thorlabs | SM1RRC | 1 | 8.57 |
| SM1 to M12 x 0.5 Lens Cell Adapter | Thorlabs | S1TM12 | 1 | 18.72 |
| Lens tube sleeve coupler | Thorlabs | SM1CPL10 | 1 | 28.28 |
| Kinematic Mirror Mount | Thorlabs | KCB1C | 1 | 110.47 |
| Ø1" Broadband Dielectric Mirror | Thorlabs | BB1-E02 | 1 | 58.01 |
| Cage Assembly Rod, 1" Long, Ø6 mm, 4 Pack | Thorlabs | ER1-P4 | 2 | 29.66 |
| N-BK7 Best Form Lens, Ø1", f = 75 mm, ARC: 350-700 nm | Thorlabs | LBF254-075-A | 1 | 43.88 |
| Cage Cube with Dichroic Filter Mount | Thorlabs | CM1-DCH | 1 | 134.77 |
| lens retaining ring | Thorlabs | SM1RR | 3 | 10.44 |
| SM1-RMS adapter | Thorlabs | SM1A3 | 1 | 13.88 |
| Lens tube coupler | Thorlabs | SM1T4 | 1 | 20.76 |
| End cap | Thorlabs | SM1CP2 | 1 | 14.48 |
| M30.5-SM1 adapter | Thorlabs | SM1A15 | 1 | 17.45 |
| M27-SM1 adapter | Thorlabs | SM1A36 | 1 | 16.4 |
| Thorlabs SM1TC lens tube clamp | Thorlabs | SM1TC | 1 | 35.15 |
| Adjustable 1.5" post mounting clamp | Thorlabs | C1511 | 1 | 56.67 |
| Condenser Lens | Thorlabs | ACL2520U-DG6-A | 1 | 23.71 |
| Cage Plate for the condenser | Thorlabs | CP33-SM1 | 2 | 14.32 |
| Cage Plate for the condenser f = 30 mm | Thorlabs | CP33T | 1 | 18.78 |
| Cage rods - 2" | Thorlabs | ER2-P4 | 2 | 35.82 |
| 90° flip mount for | Thorlabs | FM90 | 1 | 68.21 |
| breadboard 10" x 12" | Thorlabs | MB1012 | 1 | 109.8 |
| horizental 0.5" post for the LED matrix | Thorlabs | TR4 | 1 | 4.76 |
| 4-Pin Female Mating Connector for Mounted LEDs | Thorlabs | CON8ML-4 | 1 | 25.58 |
| Ø1.5" Dynamically Damped Post, 10" Long | Thorlabs | DP10A/M | 1 | 169.28 |
| Right-Angle End Clamp | Thorlabs | RA180/M | 1 | 8.81 |
| Ø12.7 mm Optical PostL = 250 mm | Thorlabs | TR250/M | 1 | 7.69 |
| Ø1/2" Optical Post L = 6" | Thorlabs | TR6 | 1 | 5.78 |
| Universal Post Holder | Thorlabs | UPH6 | 1 | 29.94 |
| Clamping Fork 5 Pack | Thorlabs | CF038-P5 | 1 | 33.42 |
| Olympus UPLFLN 40X objective | Edmund | #86-860 | 1 | 1215.5 |
| XY stage + cable | Heidstar | HDS-U-XY6060SN | 1 | 1669.70 |
| Z stage + cable | Heidstar | HDS-S-Z6SN-STF | 1 | 882.43 |
| Controller + LED driver | Heidstar | N/A | 1 | 755.04 |
| LED matrix | Heidstar | N/A | 1 | 144.26 |
| Joystick + cable | Heidstar | N/A | 1 | 700.44 |
| Monochrome camera + trigger + USB | Heidstar | MER2-630-60U3M | 1 | 205.14 |
| 50mm imaging lens | Heidstar | MVL-F5024M-10MP | 1 | 106.08 |
| AlexaFluor 488 filter cube | N/A | N/A | 1 | 425.38 |
| Studded Pedestal Base Adapter | Thorlabs | BE1/M | 4 | 32.48 |
| Pillar post m6 taps L = 50mm | Thorlabs | RS50/M | 4 | 76.84 |
| Pillar post m6 taps L = 75mm | Thorlabs | RS75/M | 4 | 83.84 |
| adapter sm1 to RMS | Thorlabs | SM1A36 | 1 | 13.88 |
| USB 3.0 A Male to A Male Lead, 1m Blue | Farnell | CAC250018 | 1 | 3.70 |
| HYLEC IP65 ENCLOSURE | Screwfix | 2597G | 1 | 31.65 |
| 36 mdr cable | Ebay | mdfly electronics | 1 | 40.92 |
| Linux laptop | Dell | Latitude 5440 | 1 | 799.9 |
| | | | **Total =** | 8645.74 |

Syringe pump and controls

| **Part** | **Vendor** | **Code** | **Quantity** | **total cost** |
|---|---|---|---|---|
| TB6600 stepper driver x6 | Robotshop | RB-Dfr-727 | 6 | £95.16 |
| PLA Filament | RS | 832-0264 | 1 | £30 |
| Linear bearings 8mm ID x10 | Robotshop | RB-Sct-1372 | 10 | £42.30 |
| 10 mm bearing | RS | 441-9969 | 5 | £25.05 |
| limit switches | Robotshop | RB-Tam-71 | 6 | £9.90 |
| Nema stepper motor x3 | Amazon | N/A | 2 | £47.98 |
| Shaft coupler | Robotshop | RB-Sct-1160 | 6 | £36.60 |
| Flexible couplings | NU Workshop | N/A | 5 | £48.72 |
| M10 threaded rod | NU Workshop | N/A | 5 | £87.86 |
| screws | Screwfix | N/A | 1 | £12.49 |
| Wago 5-Way Terminal Block, 32A, Spring Cage Terminals, 24 # 12 AWG, Cable Mount | RS | 883-7557 | 1 | £6.79 |
| RS PRO Straight, Panel Mount, Socket to Socket Type B to A 2.0 USB Connector | RS | 916-0227 | 1 | £9.67 |
| RS PRO 4 Pole Din Socket, 2A, 100 V ac, Twist Lock, Female, Panel Mount | RS | 786-3429 | 1 | £3.38 |
| HYLEC IP65 WEATHERPROOF OUTDOOR ENCLOSURE 300 X 220 X 400MM | Screwfix | 6608G | 1 | £52.84 |
| GOLDSCREW PZ COUNTERSUNK MULTIPURPOSE SCREWS 4 X 25MM 200 PACK | Screwfix | 17430 | 1 | £2.69 |
| GOLDSCREW PZ COUNTERSUNK MULTIPURPOSE SCREWS 3 X 25MM 200 PACK | Screwfix | 14448 | 1 | £1.99 |
| RS PRO Square Bracket for Top Hat DIN Rail 10PK | RS | 467-383 | 1 | £29.24 |
| Weidmuller Red WDU Feed Through Terminal Block, Single level, 2.5mm², ATEX, 800 V 10 PK | RS | 779-649 | 2 | £15.86 |
| Weidmuller Blue WDU Feed Through Terminal Block, Single level, 2.5mm², ATEX, 800 V 10PK | RS | 425-285 | 2 | £14.40 |
| Weidmuller Black WDU Feed Through Terminal Block, Single level, 2.5mm², ATEX, 800 V 10PK | RS | 779-627 | 2 | £15.86 |
| Weidmuller W ATEX End Cover for DIN Rail Terminal Blocks 5PK | RS | 425-291 | 3 | £4.53 |
| Weidmuller EW ATEX End Stop for DIN Rail Terminal Blocks | RS | 425-314 | 1 | £6.74 |
| Weidmuller WQV Jumper Bar for DIN Rail Terminal Blocks | RS | 425-336 | 3 | £47.82 |
| Weidmuller ATEX 2 Way Screw Down WPE 2.5, 60mm Length 30 → 12 AWG | RS | 193-124 | 1 | £11.74 |
| RS PRO Steel Slotted Din Rail, Top Hat Compatible, 1m x 35mm x 7.5mm | RS | 467-416 | 1 | £19.94 |
| 6 core cable 100m reel, unscreened white PVC | RS | 8200182 | 1 | £59.82 |
| Arduino mega controller | RS | 769-7412 | 1 | £35.70 |
| adjustable spanner | Screwfix | 86962 | 1 | £9.99 |
| long nut | Screwfix | 728GX | 1 | £1.79 |
| | | | **Total =** | £786.85 |

Experimental containment and fitting

| **Part** | **Vendor** | **Code** | **Quantity** | **total cost** |
|---|---|---|---|---|
| K 470 universal container, IP 65 | ZARGES | 366218 | 1 | £700 |
| Anti Vibration Mounts | WDS | WDS 719-20540 | 4 | 30.68 |
| Toggle latch | WDS | 4202-201K2 | 12 | 21.72 |
| Machined aluminium parts | N/A | N/A | 1 | 200 |
| Aluminium extrusion and brackets | N/A | N/A | 1 | 227.87 |
| Aluminium plates | metal supermarkets | N/A | 1 | 499.1 |
| Dual lock | Amazon | N/A | 1 | 14.9 |
| balloons | Amazon | N/A | 1 | 3.89 |
| Noga HSS RB 1000 Deburring Tool for Deburring | RS | 339-4336 | 1 | 12.03 |
| RS PRO IEC C7 Socket to BS 1363 UK Plug Power Cord, 1.5m | RS | 452-669 | 2 | 19.88 |
| RS PRO Pozidriv Countersunk A4 316 Stainless Steel Machine Screws DIN 965, M4x16mm | RS | 158-3629 | 1 | 9.23 |
| Plain Stainless Steel, Hex Bolt, M10 x 35mm | RS | 797-6282 | 1 | 23.39 |
| RS PRO Black, Self-Colour Steel Hex Socket Countersunk Screw, DIN 7991, M8 x 40mm | RS | 822-9149 | 1 | 11.12 |
| RS PRO Black Nylon Cable Tie, 100mm x 2.5 mm | RS | 233-455 | 2 | 5.38 |
| RS PRO Self Adhesive Black Cable Tie Mount 12.5 mm x 12.5mm, 3.2mm Max. Cable Tie Width | RS | 811-1723 | 4 | 32.2 |
| 3M Dual Lock™ SJ3540 Black Hook & Loop Tape, 25mm x 2.5m | RS | 163-4737 | 1 | 33.51 |
| 10m Magnetic Tape, Adhesive Back, 0.75mm Thickness | RS | 297-9116 | 1 | 14.96 |
| DETA RUBBER GASKETS 68MM 10 PACK (9076J) | Screwfix | 9076J | 1 | 1.43 |
| NO-NONSENSE GENERAL-PURPOSE SILICONE CLEAR 310ML | Screwfix | 35887 | 1 | 4.59 |
| silicone tubing | Amazon | N/A | 1 | 55.18 |
| Mosquito Net | N/A | N/A | 1 | 10.99 |
| Big cable ties | N/A | N/A | 1 | 7.9 |
| PLA Filament | RS | 832-0264 | 1 | £30 |
| | | | **Total =** | £1970 |